\documentclass[iop,referee]{emulateapj-rtx4}

\usepackage{graphicx}  
\usepackage{dcolumn}  
\usepackage{bm}           
\usepackage{amsfonts,amsmath,amssymb,mathrsfs}
\usepackage{color}
\usepackage{hyperref}
\hypersetup{
  colorlinks=true,        
  linkcolor=blue,         
  citecolor=cyan,         
}

\usepackage{natbib,times}
\newcommand{\cqg}{{CQGra}}
\renewcommand{\apjl}{{ApJL}}

\renewcommand{\nat}{{Natur}}

\shorttitle{X-ray flashes from long-lived neutron stars}

\shortauthors{R. Ciolfi}

\begin{document}
\title{X-ray flashes powered by the spindown of long-lived neutron stars}

\author{Riccardo Ciolfi}

\affil{Physics Department, University of Trento, Via
  Sommarive 14, I-38123 Trento, Italy\\
INFN-TIFPA, Trento Institute for Fundamental Physics
  and Applications, Via Sommarive 14, I-38123 Trento, Italy}

\email{riccardo.ciolfi@unitn.it}

\begin{abstract}

X-ray flashes (XRFs) are a class of high-energy transients whose nature is still open to question. Similar in many aspects to common gamma-ray bursts (GRBs), their strong X-ray emission is accompanied by very low or absent emission in the gamma-ray band. Despite this key difference, a number of indications have consolidated the idea that XRFs and GRBs share a common origin, including a number of potential XRF/supernova associations and the consistency of some XRFs with the Amati relation for long GRBs. However, the difficulties in explaining XRFs as off-axis or intrinsically weak GRBs still cast doubts on this interpretation.
Here we explore the possibility that some XRFs are instead powered by the spindown of a long-lived neutron star (NS) formed in a binary NS (BNS) merger or, possibly, in a core-collapse supernova.
Focusing on XRF 020903 and a few other cases observed by {\it HETE-2}, we show that their lack of gamma-ray emission, spectral properties, duration and X-ray luminosity find a natural explanation within our hypothesis. Moreover, we point out that the agreement of XRF 020903 with the Amati and Ghirlanda relations for long GRBs is respectively only marginal and problematic.
Assuming a BNS merger origin for the long-lived NS, we use XRF observations to estimate a lower limit on the rate of BNS mergers accompanied by a potentially observable XRF signal. Within the reach of the advanced LIGO and Virgo gravitational wave detectors, we find $> 0.02-0.05~\mathrm{yr}^{-1}$.
Finally, we discuss the implications of a supernova association for the XRF events considered.
\end{abstract}
		

\section{Introduction}
\label{intro}

Commonly interpreted as a subclass of gamma-ray bursts (GRBs), X-ray flashes (XRFs) are characterized by surprisingly low energy fluences in the gamma-ray band and higher fluences in the X-ray band (i.e.~$S_X>S_\gamma\,$, where ``$X$" refers to $2-30~\mathrm{keV}$ and ``$\gamma$" to $30-400~\mathrm{keV}$). Since their discovery with {\it BeppoSAX} 
\citep{BeppoSAX} and the numerous observations by {\it HETE-2} \citep{Ricker2003,Sakamoto2005}, the interpretation of XRFs has been uncertain. 
Apart from having a much lower peak photon energy $E_\mathrm{p}$, their spectral properties appear compatible with those of typical GRBs. Moreover, a careful analysis of an event with known redshift, XRF 020903, has revealed its consistency with the $E_\mathrm{p} -E_\mathrm{iso}$ relation found for long GRBs \citep{Amati2002}, where $E_\mathrm{iso}$ is the isotropic energy emission in the $1-10,000~\mathrm{keV}$ band. This collective evidence led to the conclusion that XRFs are GRBs \citep{XRF020903}. 
This view is further supported by a number of putative XRF/supernova (SN) associations (e.g.~\citealt{Soderberg2005,Pian2006}) and the existence of ``X-ray-rich'' GRBs, partially filling the gap between typical GRBs and XRFs \citep{Sakamoto2005}. 
Nevertheless, current models of XRFs within the GRB paradigm encounter difficulties in explaining their softer emission, casting doubts on their nature. 
Possible explanations include GRBs observed off-axis (\citealt{Yamazaki2002,Zhang2004,Urata2015}) or inefficient/subenergetic GRB jets \citep{Huang2002}. The possibility that XRFs are far distant GRBs highly redshifted to lower photon energies has been generically ruled out (e.g.,~\citealt{XRF020903}).

In this paper, we propose an alternative explanation for XRFs, independent of GRBs and based on the emission powered by the spindown of a long-lived neutron star (NS) formed as the end product of either a binary NS (BNS) merger or the death of a massive star.
We focus our analysis on XRF 020903 and a few other XRFs observed by {\it HETE-2} with no significant gamma-ray emission ($S_X/S_\gamma>5$) to show that our hypothesis is consistent with the observations and that it offers a natural explanation for the puzzling properties of these events.

\section{X-ray emission from a long-lived NS}
\label{EM-spindown}

Since the observation of NSs with a mass of $\sim\,$2~$\mathrm{M}_\odot$
(\citealt{Demorest2010,Antoniadis2013}) it has been clear that the maximum mass of uniformly rotating NSs is $\gtrsim 2.4~\mathrm{M}_\odot$ \citep{Lasota1996}. Moreover, the distribution of masses in merging BNSs favors the combination $\sim\,$$1.3-1.4$~$\mathrm{M}_\odot$ \citep{Belczynski2008a}, suggesting a typical mass of the merger product of $\sim\,$$2.3-2.4~\mathrm{M}_\odot$.
This leads to the conclusion that a significant fraction of BNS mergers should result in long-lived NSs, namely NSs that can either survive for about a spindown timescale before eventually collapsing to a black hole (BH) or NSs that will never collapse. 
The spindown of a long-lived NS can power strong and sustained electromagnetic emission \citep{Yu2013, Metzger2014}, which has been invoked to explain the long-lasting ($\sim\,$$100-10^{5}$~s) X-ray afterglows observed by {\it Swift} \citep{Swift} in association with many short GRBs (\citealt{Zhang2001,Metzger2008a,Rowlinson2013,Gompertz2013,Gompertz2014,Lu2015}).  

Recently, \cite{Siegel2016a,Siegel2016b} proposed a detailed model to describe the evolution of a long-lived NS and its electromagnetic emission 
up to $\sim \!10^{7}$~s, taking into account the crucial role of the matter ejected via post-merger baryon-loaded winds.
Exploring a wide range of physical parameters, they found that the spindown-powered signal has a delayed onset ($\sim\,$$10-100$~s) and peaks $\sim100-10^{4}$~s after merger. 
Furthermore, it typically falls inside the X-ray band $0.5-10~\mathrm{keV}$ and the maximum luminosity is $\sim10^{46}-10^{49}~\mathrm{erg\, s}^{-1}$. 
Additionally, the emission is highly isotropic.
Concerning the spectral properties, gamma-ray emission is negligible and the first part of the signal is expected to be predominantly thermal, with black body (BB) temperatures of $\sim0.1-\mathrm{few}~\mathrm{keV}$ at maximum luminosity.
After the maximum, the BB temperature can rapidly decrease (factor 2 in $\sim10-10^{3}$~s).
Days/weeks after merger, the emission gradually shifts to lower energies and an optical transient can be produced, with luminosities potentially comparable to core-collapse supernovae (SNe) (cf. Figure 1 of \citealt{Siegel2016b}).
BNS mergers are among the most promising sources of gravitational waves (GWs) \citep{Abadie2010} and given the high occurrence, isotropy, high luminosities and long duration, \citet{Siegel2016a,Siegel2016b} concluded that these spindown-powered signals represent very promising counterparts to the GW signals from such mergers. 

As a note of caution, the model of \citet{Siegel2016a,Siegel2016b} employs a simplified treatment for the dynamics of the ejected matter surrounding the NS in which the ejecta shell remains constant in thickness. The internal pressure of the ejecta that would tend to increase the thickness of this shell could be partially compensated by the high pressure of the photon-pair plasma nebula confined inside, although further investigation and more refined modeling is needed to shed light on this aspect. Within a different treatment, envisaging an increase of the ejecta shell thickness, the parameter space for which the emission falls in the X-ray band is likely to be reduced.
\footnote{For example, if this shell is allowed to expand with the local sound speed, in order to have X-ray emission the rotational energy of the remnant has to be $\gtrsim 10^{52}$~erg, the dipolar magnetic field strength $\gtrsim 10^{15}$~G, and the total mass in the isotropic post-merger ejecta $\lesssim 10^{-3}$~M$_\odot$. 
While rotational energies as high are typically found in BNS merger simulations, it is not clear how strong the external dipolar magnetic field can be when the remnant has settled to the spindown regime (within seconds after merger), even knowing that the total magnetic field strength inside the remnant can easily reach $10^{15}-10^{16}$~G.
The mass launched by isotropic winds from the remnant is expected to be a very small fraction of a solar mass, but the outcome is very uncertain and can change significantly for different BNS systems. In particular, in some cases it is expected to exceed $\sim10^{-3}$~M$_\odot$. 
When higher mass and weaker magnetic fields are considered, the emission shifts to lower energy bands and the model predicts a signal that falls out of the X-ray band, as in \citet{Yu2013}.}

Although the model presented in \citet{Siegel2016a,Siegel2016b} is focused on long-lived NSs formed in BNS mergers, a similar emission might be expected if the NS is formed in association with a SN, 
provided that at peak luminosity ($\sim100-10^4~\mathrm{s}$ since NS formation) the exploding outer envelope of the progenitor star has been efficiently removed along the direction of the observer by, e.g., the early magnetized outflow from the proto-NS (\citealt{Bucciantini2008,Bucciantini2009}).
In this case, the observer should be roughly on-axis with the NS spin and the early outflow should not be powerful enough to generate a GRB (see also Section~\ref{sec:conclusion}). 

\begin{table*}[tbh]
\centering
\caption{XRF sample properties: duration, peak photon energy ($E_\mathrm{p}$), X-to-gamma fluence ratio ($S_X/S_\gamma$), peak photon number flux ($F^N_\mathrm{p}$), redshift ($z$), black-body temperature ($T_\mathrm{BB}$), fluence ratio recomputed assuming thermal spectrum ($S_X/S_\gamma~(\mathrm{BB})$).}
\label{table}
{\renewcommand{\arraystretch}{1.6}
\begin{tabular}{cccccccp{0.1\textwidth}}
\hline\hline
XRF & Duration~[s]\footnotemark[1] & $E_\mathrm{p}$~[keV]\footnotemark[1] & $S_X/S_\gamma$\footnotemark[1] & $F^N_\mathrm{p}$ ($2-30$ keV) ~[cm$^{-2}$ s$^{-1}$]\footnotemark[1] & $z$\footnotemark[2] & $T_\mathrm{BB}$~[keV]\footnotemark[3] & $S_X/S_\gamma~(\mathrm{BB})$ \\
\hline
020903 & 13.00 & $2.6^{+1.4}_{-0.8}$ & 7.31 & $2.75 \pm 0.66$ & $0.25\pm0.01$ & $0.68-1.3$ & $\geq 4.4 \times 10^6$ \\
\hline
010213 & 34.41 & $3.41^{+0.35}_{-0.40}$ & 11.38 & $6.33 \pm 0.77$ & - & $1.2$ & $2.1 \times 10^7$ \\
011130 & 50.00 & $<3.9$ & 5.96 & $5.27 \pm 1.27$ & - & $<1.4$ & $> 1.3 \times 10^6$ \\
020625 & 41.94 & $8.52^{+5.44}_{-2.91}$ & 20.49 & $2.86 \pm 0.97$ & - & $3.0$ & $97$ \\
030723 & 31.25 & $<8.9$ & 7.47 & $1.98 \pm 0.38$ & - & $<3.2$ & $> 72$ \\
\hline
\end{tabular}
\footnotetext[1]{From \citet{Sakamoto2005}. Obtained assuming Band function, power-law, or power-law plus exponential cutoff spectrum.}
\footnotetext[2]{From \citet{Soderberg2004}.}
\footnotetext[3]{For XRF 020903 we employ BB temperature estimates by \citet{XRF020903} (error $\lesssim40\%$) and apply redshift corrections. For the other cases we take $E_\mathrm{p}$ as the peak photon energy of the BB spectrum and compute the corresponding BB temperature (no redshift correction).} 
}
\end{table*}

\section{XRF 020903}
\label{020903}

In order to link XRFs to the spindown-powered X-ray emission from long-lived NSs, we consider cases with $S_X/S_\gamma>5$ and no significant gamma-ray emission. This choice is more conservative than the common definition of XRFs  ($S_X/S_\gamma>1$) and allows us to avoid contamination from ``X-ray-rich'' GRBs. 
Here we focus on the best studied event with known redshift, XRF 020903, while in the next Section we extend the discussion to a few more cases. 

The main properties of XRF 020903 are summarized in Table \ref{table}. 
With no significant detection of photons above 10~keV, this event represents an ideal candidate for our analysis. 
In addition to the lack of gamma-ray emission, another strikingly different property of this event compared to typical GRBs is that data can be fit with a mostly thermal spectrum, with an initial BB temperature of $\sim\,$1~keV, decreasing to $\sim\,$0.5~keV within $\sim\,$10~s \citep{XRF020903}. 
These properties meet the expectations of our model. 
Note that without fluence observed above 30 keV, $S_\gamma$ depends entirely on the assumed spectral behavior at high energies. 
The fluence ratio estimated with a power-law spectral fit is $S_X/S_\gamma\simeq7$ \citep{Sakamoto2005}, while adopting a pure thermal spectrum would give $S_X/S_\gamma\gtrsim7\times 10^8$.

The X-ray signal emerges above the detector noise for $\sim$13~s \citep{XRF020903} and from a rough estimate based on photon counts the noise level is at $\sim\,$80\% of the maximum luminosity. From the parameter study by \citet{Siegel2016b} the corresponding signal duration, which is mostly controlled by the mass ejected via NS winds $M_\mathrm{wind}\sim10^{-4}-10^{-1}~\mathrm{M}_\odot$, is $\sim100-10^{3}$~s. 
A shorter duration of $\sim\,$10~s is still compatible with the model, but suggests $M_\mathrm{wind}\lesssim 10^{-5}~\mathrm{M}_\odot$.

Observations of XRF 020903 days/weeks after the trigger revealed an optical signal, from which it was possible to place the event at redshift $z=0.25\pm0.01$ \citep{Soderberg2004}. This excludes that XRF 020903 is an ordinary GRB at large distance (i.e.~$z\sim 100$, \citealt{XRF020903}). 
Applying the redshift correction to the BB temperature, the X-to-gamma fluence ratio for pure thermal spectrum would become $S_X/S_\gamma\gtrsim4\times 10^6$. 

The BNS merger scenario holds as long as the optical signal can be explained in terms of late-time spindown-powered emission.\footnote{
According to \citet{Siegel2016b}, optical luminosities of $\simeq 10^{42}-10^{43}~\mathrm{erg\, s}^{-1}$ can be expected $1-10$ days after merger, with a lightcurve possibly similar to a SN.
In principle, also the radio signal observed $30-300$ days after the trigger \citep{Soderberg2004} could be explained as spindown-powered emission, although this conclusion is more speculative.} 
Conversely, a SN \citep{Soderberg2005,Bersier2006} would imply a long-lived NS formed from the death of a massive star (see Section~\ref{sec:conclusion}).

Assuming a standard $\Lambda \mathrm{CDM}$ cosmology with the parameters reported in \cite{Planck2015}, we estimate a luminosity distance of $d_L\sim1.3~\mathrm{Gpc}$. 
From the peak flux of $\simeq\,$14.7$\times 10^{-9}~\mathrm{erg}\,\mathrm{s}^{-1}\mathrm{cm}^{-2}$ ($\sim\,$35\% error) in the $2-10$~keV band \citep{XRF020903}, the resulting intrinsic luminosity in the $2.5-12.5$~keV band is $\simeq 3\times 10^{48}~\mathrm{erg\, s}^{-1}$.
This value is consistent with the expected range of X-ray luminosities for a long-lived NS\footnote{Since the maximum luminosity depends mostly on the initial spindown luminosity of the NS, we can refer to the results of \citet{Siegel2016b} to roughly estimate the corresponding dipolar magnetic field strength $B_\mathrm{p}$, for a typical initial rotational energy of $E_\mathrm{rot}\sim \mathrm{few}\times10^{52}$~$\mathrm{erg}$. A peak luminosity of $\sim10^{48}$~$\mathrm{erg\, s}^{-1}$ gives $B_\mathrm{p}\sim 10^{15}$~$\mathrm{G}$.}
\citep{Siegel2016b}. 

We conclude that the observed properties of XRF 020903 are consistent with the spindown-powered emission from a long-lived NS formed in a BNS merger or, possibly, in connection to a SN. In particular, the lack of gamma-ray emission and the compatibility with a predominantly thermal spectrum with BB temperatues of $\sim\,$1~keV are naturally explained. These properties, at the same time, represent a challenge for any attempt to model XRF 020903 as a GRB.

 \section{A sample of candidate XRF events}
\label{sample}

\begin{figure}
  \includegraphics[width=0.465 \textwidth]{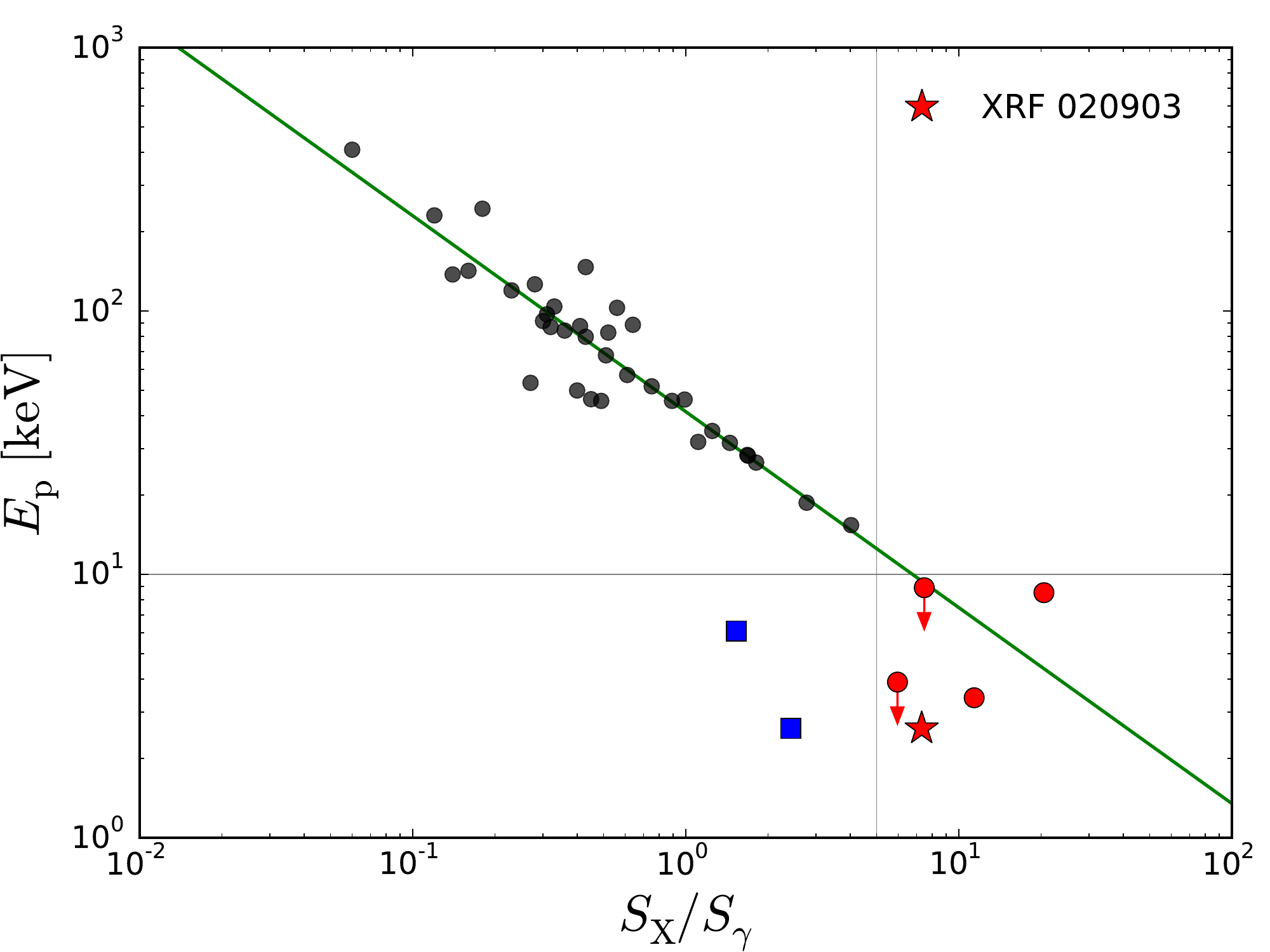}
  \includegraphics[width=0.465 \textwidth]{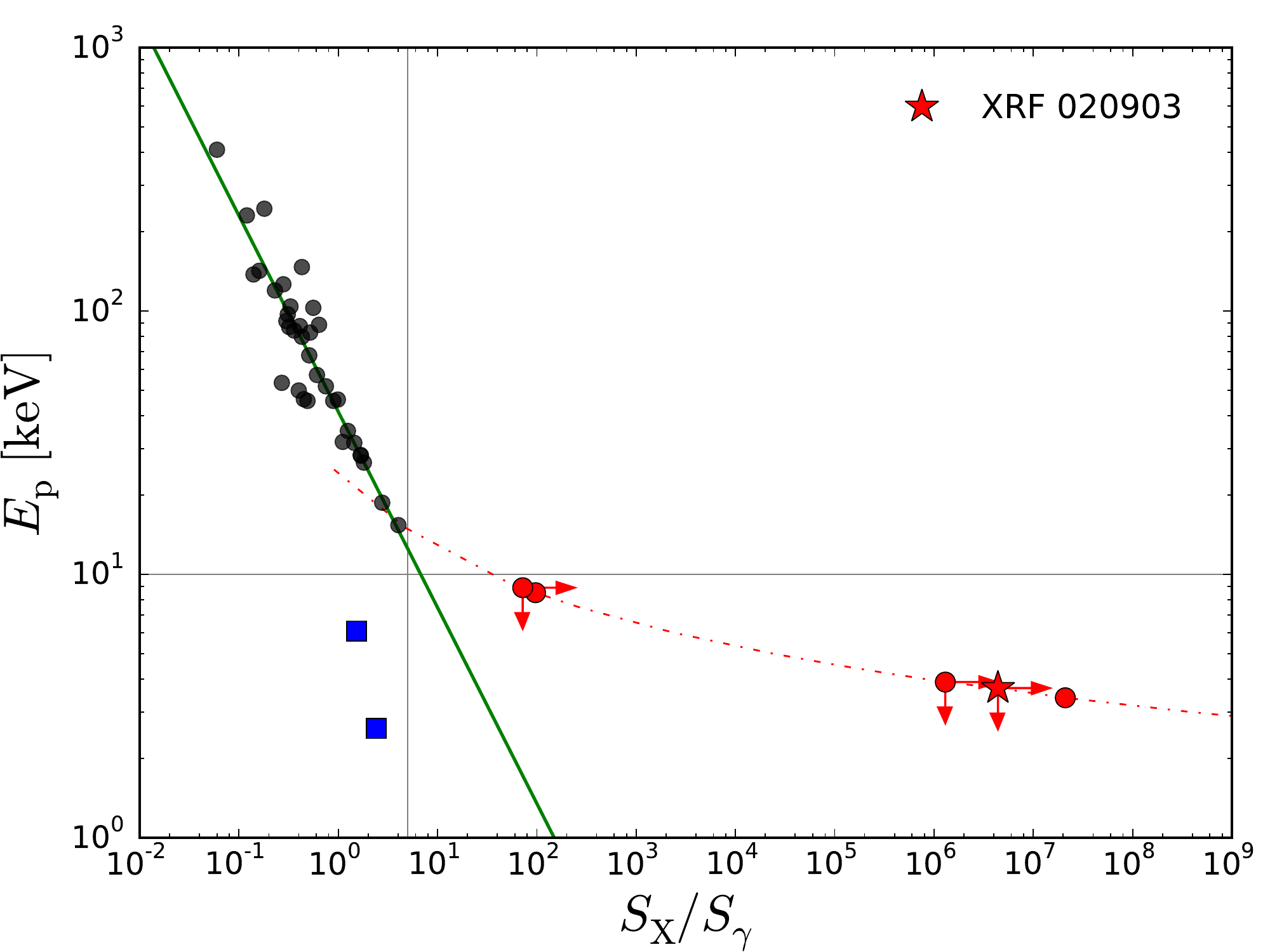}
  \caption{Top: peak photon energy vs. X-to-gamma fluence ratio for the events in the sample of \citet{Sakamoto2005}. Errors are not shown and no redshift corrections are applied. Our selected events are in red (star and large circles). Blue squares are cases with $E_\mathrm{p}<10$~keV and $1<S_X/S_\gamma<5$ (not included in our analysis). The green line indicates the correlation $E_\mathrm{p}\propto(S_X/S_\gamma)^{-0.74}$ (see text). 
Bottom: X-to-gamma fluence ratio recomputed with the assumption of pure thermal emission for the five selected cases. XRF 020903 is also corrected for redshift, resulting in  $T_\mathrm{BB}\lesssim1.3$~keV. The red dotted-dashed line shows the profile for a BB spectrum.}
\label{fig1}
\end{figure}

Here we extend our discussion to all XRFs with $S_X/S_\gamma>5$ in the sample of \citet{Sakamoto2005}, where the X-to-gamma fluence ratios were obtained under the assumption of power-law (with/without exponential cutoff) or Band function spectrum. The properties of these events (five including XRF 020903) are given in Table~\ref{table}.
Figure~\ref{fig1} (upper panel) shows the peak photon energy $E_\mathrm{p}$ versus $S_X/S_\gamma$ for the entire sample of \citet{Sakamoto2005}. 
Above $E_\mathrm{p}\simeq 10~\mathrm{keV}$, there is a tight correlation given by $E_\mathrm{p}~[\mathrm{keV}]=b (S_X/S_\gamma)^{a}$, where $a\simeq -0.74$ and $b\simeq 41$.
Our selected cases all have $E_\mathrm{p}<10~\mathrm{keV}$ and appear broadly consistent with the above correlation, although with a larger scatter. 
Two more events have $E_\mathrm{p}<10~\mathrm{keV}$, but $1<S_X/S_\gamma<5$. Although they are potential candidates, we exclude them from our analysis.

As for XRF 020903, all the events in our sample lack a significant gamma-ray emission. 
Since we cannot rely on a detailed spectral analysis like the one presented for XRF 020903 in \citet{XRF020903}, we proceed under the assumption that they can be described by a dominant BB spectrum with peak photon energy approximately given by the value of $E_\mathrm{p}$ estimated by \citet{Sakamoto2005}, and compute the corresponding BB temperature.\footnote{A mainly thermal spectrum with high-energy non-thermal tail is probably necessary to fit the data, in which case a description in terms of a purely thermal spectrum is oversimplified.} We cannot apply redshift corrections for these events. Nevertheless, if their distance is comparable to XRF 020903, corrections are small. 
Recomputing the X-to-gamma fluence ratio for a purely thermal spectrum, we find that our sample would have $S_X/S_\gamma>72$ (see Table~\ref{table}).
In the absence of a significant detection of photons above 30 keV, the consistency with the correlation satisfied by all other events with $E_\mathrm{p}>10~\mathrm{keV}$ depends entirely on the assumption of Band function or power-law spectrum.
Assuming a thermal spectrum gives a completely different result (see the lower panel of Figure~\ref{fig1}). 

The event durations in our sample are quite homogeneous with a range $13-50$~s (average $\simeq 34$~s) and compatible with the hypothesis of spindown-powered emission from a long-lived NS.
Since the redshift is known only for XRF 020903, we cannot infer the instrinsic luminosity of the other cases. 
The peak photon number flux in the $2-30$ keV band is also very homogeneous, differing only by a factor of $\simeq0.7-2.3$ from the one measured for XRF 020903 (see Table~\ref{table}). By rescaling the X-ray fluence with this factor to obtain a rough estimate we conclude that: (i) if all five events were at the same distance ($\simeq\,$1.3~Gpc), the range of intrinsic X-ray luminosities (in the $2.5-12.5$~keV band) would be $\sim (2-7)\times 10^{48}~\mathrm{erg\, s}^{-1}$; (ii) if all five events had the same X-ray luminosity, the range of distances would be  $\sim 0.9-1.5$~Gpc. 

From the properties discussed above, we conclude that a tentative extension of our interpretation of XRF 020903 to the other four cases considered is compatible with the observations.

 \section{Amati and Ghirlanda relations}
\label{sec:amati} 

\begin{figure}
  \includegraphics[width=0.465 \textwidth]{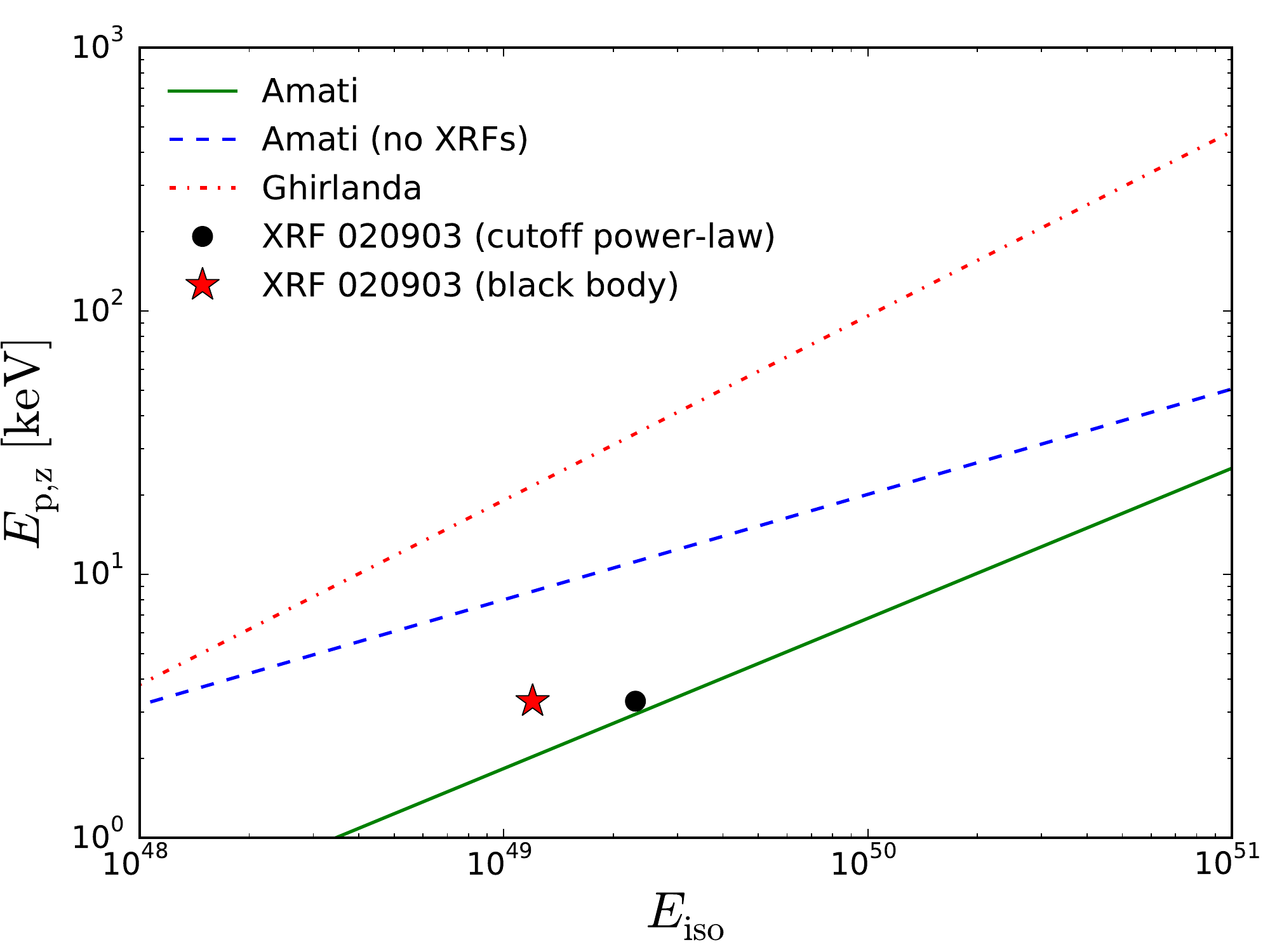}
  \caption{$E_\mathrm{p,z}-E_\mathrm{iso}$ relation for XRF 020903 assuming cutoff power-law spectrum (black dot) or thermal spectrum (red star). Continuous green line and dashed blue line correspond to the Amati relation incluing or excluding XRFs. Dot-dashed red line corresponds to the Ghirlanda relation (see text).}
\label{fig2}
\end{figure}

Assuming a cutoff power-law spectrum, \citet{XRF020903} estimated for XRF 020903 an isotropic energy emission in the source rest frame of $E_\mathrm{iso} \simeq 2.3 \times 10^{49}$~erg in the canonical band $(1-10,000)$~keV. This value, combined with the estimated peak photon energy in the source rest frame $E_\mathrm{p,z}\simeq 2.6~(1+z)\simeq 3.3$~keV, is roughly consistent with the
$E_\mathrm{p,z} -E_\mathrm{iso}$ relation found by \citet{Amati2002} for long GRBs.
This result was taken as a strong indication in favor of the interpretation of XRFs as GRBs \citep{XRF020903}.

A more recent version of the Amati relation, including XRF 020903, gives
$E_\mathrm{p,z}~\mathrm{[keV]}\simeq94 \times (E_\mathrm{iso}/10^{52}~\mathrm{erg})^{0.57}$ (\citealt{Amati2008}; see Figure~\ref{fig2}). As noted by \citet{Ghirlanda2004}, however, 
the slope of the Amati relation depends on whether XRFs are included or not in the fit. Excluding XRFs, \citet{Ghirlanda2004} found a different relation where $E_\mathrm{p,z}\propto E_\mathrm{iso}^{\,0.4}$ (see Figure~\ref{fig2}), which is only marginally consistent with XRF 020903 (factor $\sim\,$20 discrepancy in $E_\mathrm{iso}$).
This shows that the Amati relation cannot be taken as compelling evidence that XRFs and GRBs are the same phenomenon.

Taking into account the opening angle of a sample of GRBs (as inferred from the jet break in the lightcurve) \citet{Ghirlanda2004} found the relation $E_\mathrm{p,z}~\mathrm{[keV]}\simeq480 \times (E_\theta/10^{51}~\mathrm{erg})^{0.7}$, where $E_\theta=(1-\cos{\theta}) E_\mathrm{iso}$ is the collimation-corrected energy emission and $\theta$ the half opening angle. In order to be reconciled with the Ghirlanda relation, XRF 020903 should have $\theta\sim 25^\mathrm{o}$ (\citealt{Ghirlanda2004}; see also Figure \ref{fig2}), which is much larger than the typical range for long GRBs,
$\sim(2-10)^\mathrm{o}$.
Moreover, this opening angle would correspond to a jet brake at $\sim\,$3 days that, according to \citet{Soderberg2004}, was not observed. This would suggest that XRF 020903 is an outlier of the Ghirlanda relation and therefore its interpretation as a GRB is problematic \citep{Ghirlanda2004}.

If we interpret XRF 020903 as predominantly thermal radiation described by a BB spectrum with $E_\mathrm{p,z}\simeq 3.3$~keV and scale the total energy fluence in the $(1-10,000)$~keV band so that the observed fluence in the $2-10$~keV band ($\sim5.9\times 10^{-8}~\mathrm{erg\,cm}^{-2}$, \citealt{XRF020903}) is reproduced, we find 
$E_\mathrm{iso} \simeq 1.2\times 10^{49}$~erg. 
Unlike the collimated emission observed in many GRBs, the emission mechanism we propose for XRF 020903 is isotropic, i.e.~$\theta=\pi/2$ and $E_\mathrm{iso}=E_\theta$. This allows for a direct comparison with both the Amati and the Ghirlanda relations (see Figure \ref{fig2}). 
According to our interpration, however, the event is not a GRB and we do not have to expect an agreement with these relations. 

 \section{Concluding remarks}
\label{sec:conclusion} 

In this paper we have shown that the high-energy emission properties of XRF 020903 can be explained in terms of spindown-powered emission from a long-lived NS formed in a BNS merger or, possibly, in connection to a SN. These include spectral properties (lack of gamma-ray emission, mainly thermal spectrum, BB temperature and its evolution) and lightcurve properties (duration, luminosity). 
At the same time, these properties challenge the common interpretation of XRF 020903 as a GRB. Moreover, we have argued that the concordance with the Amati relation is not as compelling as commonly assumed and we have pointed out difficulties in reconciling the event with the Ghirlanda relation.
In addition to XRF 020903, we have selected a few other cases in the sample of \citet{Sakamoto2005}, and we have shown that our interpretation can be extended to these events with no contradiction with the observations. 
If our scenario is correct, it still does not apply to ``XRFs'' with a significant gamma-ray emission. We propose a redefinition of those events as ``X-ray-rich'' GRBs, keeping ``XRF'' only for non-GRB events.

In our discussion we have only considered events observed by {\it HETE-2} during $\sim\,$3 years of operation (2001-2003).
In this time period, any event like XRF 020903 would have been possibly detected with the wide field X-ray monitor (WXM) on board {\it HETE-2}. 
If XRF 020903 was powered by a long-lived NS formed in a BNS merger and accounting for the time of operation and field of view of WXM (e.g. \citealt{Pelangeon2008}), the detection of this event would imply a lower limit on the rate of BNS mergers leading to a long-lived NS of $> 2.8~\mathrm{Gpc}^{-3}\mathrm{yr}^{-1}$.
Rescaling with the reach of the advanced ground-based GW detectors LIGO and Virgo \citep{Harry2010,Accadia2011}, $d_\mathrm{L}\simeq 445/2.26 \simeq 197~\mathrm{Mpc}$ \citep{Abadie2010}, we obtain a rate of such merger events detectable in GWs of $> 0.02~\mathrm{yr}^{-1}$.
As a reference, the lower limit estimated in \citet{Abadie2010} gives $> f\times 0.4~\mathrm{yr}^{-1}$, where $f$ is the fraction of BNS mergers forming a long-lived NS. 
Extending to a total of four events in our selected sample (excluding XRF 020625, which was not first triggered by WXM) and assuming that they all occurred within a luminosity distance of $1.5~\mathrm{Gpc}$ (obtained by imposing that the faintest event had the same intrinsic luminosity of XRF 020903), we obtain an event rate detectable in GWs of $> 0.05~\mathrm{yr}^{-1}$. 
If forming a long-lived NS is rare ($f\ll 10\%$), the above limit is more stringent than the one provided in \citet{Abadie2010}.

If  short GRBs are produced by BNS mergers with a long-lived remnant (as, e.g., in the ``time-reversal'' scenario, \citealt{Ciolfi2015}), they should be accompanied by the isotropic X-ray signal that we propose here as an explanation for events like XRF 020903. In this case, we predict a rate of detectable short GRBs within the reach of LIGO/Virgo for face-on mergers, $d_\mathrm{L}\simeq 197\times 1.5 \simeq 296~\mathrm{Mpc}$, of $> f_\gamma f_\theta \times 0.17~\mathrm{yr}^{-1}$, where $f_\gamma$ is the fraction of mergers producing short GRBs and $f_\theta=1-\cos{\theta_j}$, with $\theta_{j}$ the typical half opening angle.
From observed short GRBs with known redshift, \citet{Metzger2012} estimate a rate of $\sim0.03~\mathrm{yr}^{-1}$.
Our lower limit is smaller by a factor $\sim6\times f_\gamma f_\theta <1$, which would suggest that our rates are pessimistic.

If all/some of the XRF events considered are instead associated with long-lived NSs born in SN explosions, the agreement with the predictions of \citet{Siegel2016a,Siegel2016b} for the BNS merger case is remarkable.  
This similarity might be explained in a scenario in which the outer envelope of the progenitor star has been removed from the line of sight of the (on-axis) observer by an early magnetized wind from the proto-NS (\citealt{Bucciantini2008,Bucciantini2009}) that was not powerful enough to produce a GRB. However, further investigation is necessary to assess the viability of this scenario. 

Finally, if XRF 020903 is associated with a SN and not powered by the spindown of a long-lived NS, its compatibility in the high-energy emission remains to be explained and might provide important hints on the nature of this type of event.
Moreover, our interpretation might still apply to any analogous event without confirmed SN association.

To clarify the nature of XRFs and exploit their potential, new X-ray missions monitoring the sky with large field of view and increased sensitivity are urgently needed. 

\acknowledgements
\noindent We thank M.~Branchesi, G.~Greco, and G.~Stratta for inspiring discussions and G.~Ghirlanda, D.M.~Siegel, and N.~Bucciantini for valuable comments. This work is supported by MIUR FIR Grant No.~RBFR13QJYF. 


\bibliographystyle{aasjournal}

\begin{thebibliography}{}
\expandafter\ifx\csname natexlab\endcsname\relax\def\natexlab#1{#1}\fi

\bibitem[{{Abadie} {et~al.}(2010){Abadie}, {Abbott}, {Abbott}, {Abernathy},
  {Accadia}, {Acernese}, {Adams}, {Adhikari}, {Ajith}, {Allen}, \&
  et~al.}]{Abadie2010}
{Abadie}, J., {Abbott}, B.~P., {Abbott}, R., {et~al.} 2010, \cqg, 27, 173001

\bibitem[{{Accadia} {et~al.}(2011){Accadia}, {Acernese}, {Antonucci}, {Astone},
  {Ballardin}, {Barone}, {Barsuglia}, {Basti}, {Bauer}, {Bebronne}, {Beker},
  {Belletoile}, {Birindelli}, {Bitossi}, {Bizouard}, {Blom}, {Bondu},
  {Bonelli}, {Bonnand}, {Boschi}, {Bosi}, {Bouhou}, {Braccini}, {Bradaschia},
  {Branchesi}, {Briant}, {Brillet}, {Brisson}, {Budzy{\'n}ski}, {Bulik},
  {Bulten}, {Buskulic}, {Buy}, {Cagnoli}, {Calloni}, {Canuel}, {Carbognani},
  {Cavalier}, {Cavalieri}, {Cella}, {Cesarini}, {Chaibi}, {Chassande Mottin},
  {Chincarini}, {Cleva}, {Coccia}, {Cohadon}, {Colacino}, {Colas}, {Colla},
  {Colombini}, {Corsi}, {Coulon}, {Cuoco}, {D'Antonio}, {Dattilo}, {Davier},
  {Day}, {De Rosa}, {Debreczeni}, {Del Pozzo}, {del Prete}, {Di Fiore}, {Di
  Lieto}, {Emilio}, {Di Virgilio}, {Dietz}, {Drago}, {Fafone}, {Ferrante},
  {Fidecaro}, {Fiori}, {Flaminio}, {Forte}, {Fournier}, {Franc}, {Frasca},
  {Frasconi}, {Galimberti}, {Gammaitoni}, {Garufi}, {G{\'a}sp{\'a}r}, {Gemme},
  {Genin}, {Gennai}, {Giazotto}, {Gouaty}, {Granata}, {Greverie}, {Guidi},
  {Hayau}, {Heidmann}, {Heitmann}, {Hello}, {Huet}, {Jaranowski}, {Kowalska},
  {Kr{\'o}lak}, {Leroy}, {Letendre}, {Li}, {Liguori}, {Lorenzini}, {Loriette},
  {Losurdo}, {Majorana}, {Maksimovic}, {Man}, {Mantovani}, {Marchesoni},
  {Marion}, {Marque}, {Martelli}, {Masserot}, {Michel}, {Milano}, {Minenkov},
  {Mohan}, {Morgado}, {Morgia}, {Mosca}, {Moscatelli}, {Mours}, {Nocera},
  {Pagliaroli}, {Palladino}, {Palomba}, {Paoletti}, {Parisi}, {Pasqualetti},
  {Passaquieti}, {Passuello}, {Persichetti}, {Pichot}, {Piergiovanni},
  {Pietka}, {Pinard}, {Poggiani}, {Prato}, {Prodi}, {Punturo}, {Puppo},
  {Rabeling}, {R{\'a}cz}, {Rapagnani}, {Re}, {Regimbau}, {Ricci}, {Robinet},
  {Rocchi}, {Rolland}, {Romano}, {Rosi{\'n}ska}, {Ruggi}, {Sassolas},
  {Sentenac}, {Sperandio}, {Sturani}, {Swinkels}, {Tacca}, {Taffarello},
  {Toncelli}, {Tonelli}, {Torre}, {Tournefier}, {Travasso}, {Vajente}, {van den
  Brand}, {Van Den Broeck}, {van der Putten}, {Vasuth}, {Vavoulidis},
  {Vedovato}, {Verkindt}, {Vetrano}, {Vicer{\'e}}, {Vinet}, {Vitale}, {Vocca},
  {Ward}, {Was}, {Yvert}, \& {Zendri}}]{Accadia2011}
{Accadia}, T., {Acernese}, F., {Antonucci}, F., {et~al.} 2011, \cqg, 28, 114002

\bibitem[{{Amati} {et~al.}(2008){Amati}, {Guidorzi}, {Frontera}, {Della Valle},
  {Finelli}, {Landi}, \& {Montanari}}]{Amati2008}
{Amati}, L., {Guidorzi}, C., {Frontera}, F., {et~al.} 2008, \mnras, 391, 577

\bibitem[{{Amati} {et~al.}(2002){Amati}, {Frontera}, {Tavani}, {in't Zand},
  {Antonelli}, {Costa}, {Feroci}, {Guidorzi}, {Heise}, {Masetti}, {Montanari},
  {Nicastro}, {Palazzi}, {Pian}, {Piro}, \& {Soffitta}}]{Amati2002}
{Amati}, L., {Frontera}, F., {Tavani}, M., {et~al.} 2002, \aap, 390, 81

\bibitem[{{Antoniadis} {et~al.}(2013){Antoniadis}, {Freire}, {Wex}, {Tauris},
  {Lynch}, {van Kerkwijk}, {Kramer}, {Bassa}, {Dhillon}, {Driebe}, {Hessels},
  {Kaspi}, {Kondratiev}, {Langer}, {Marsh}, {McLaughlin}, {Pennucci}, {Ransom},
  {Stairs}, {van Leeuwen}, {Verbiest}, \& {Whelan}}]{Antoniadis2013}
{Antoniadis}, J., {Freire}, P.~C.~C., {Wex}, N., {et~al.} 2013, Science, 340,
  448

\bibitem[{{Belczynski} {et~al.}(2008){Belczynski}, {O'Shaughnessy}, {Kalogera},
  {Rasio}, {Taam}, \& {Bulik}}]{Belczynski2008a}
{Belczynski}, K., {O'Shaughnessy}, R., {Kalogera}, V., {et~al.} 2008, \apjl,
  680, L129

\bibitem[{{Bersier} {et~al.}(2006){Bersier}, {Fruchter}, {Strolger},
  {Gorosabel}, {Levan}, {Burud}, {Rhoads}, {Becker}, {Cassan}, {Chornock},
  {Covino}, {de Jong}, {Dominis}, {Filippenko}, {Hjorth}, {Holmberg},
  {Malesani}, {Mobasher}, {Olsen}, {Stefanon}, {Castro Cer{\'o}n}, {Fynbo},
  {Holland}, {Kouveliotou}, {Pedersen}, {Tanvir}, \& {Woosley}}]{Bersier2006}
{Bersier}, D., {Fruchter}, A.~S., {Strolger}, L.-G., {et~al.} 2006, \apj, 643,
  284

\bibitem[{{Boella} {et~al.}(1997){Boella}, {Butler}, {Perola}, {Piro},
  {Scarsi}, \& {Bleeker}}]{BeppoSAX}
{Boella}, G., {Butler}, R.~C., {Perola}, G.~C., {et~al.} 1997, \aaps, 122,
  doi:10.1051/aas:1997136

\bibitem[{{Bucciantini} {et~al.}(2008){Bucciantini}, {Quataert}, {Arons},
  {Metzger}, \& {Thompson}}]{Bucciantini2008}
{Bucciantini}, N., {Quataert}, E., {Arons}, J., {Metzger}, B.~D., \&
  {Thompson}, T.~A. 2008, \mnras, 383, L25

\bibitem[{{Bucciantini} {et~al.}(2009){Bucciantini}, {Quataert}, {Metzger},
  {Thompson}, {Arons}, \& {Del Zanna}}]{Bucciantini2009}
{Bucciantini}, N., {Quataert}, E., {Metzger}, B.~D., {et~al.} 2009, \mnras,
  396, 2038

\bibitem[{{Ciolfi} \& {Siegel}(2015)}]{Ciolfi2015}
{Ciolfi}, R., \& {Siegel}, D.~M. 2015, \apjl, 798, L36

\bibitem[{{Demorest} {et~al.}(2010){Demorest}, {Pennucci}, {Ransom}, {Roberts},
  \& {Hessels}}]{Demorest2010}
{Demorest}, P.~B., {Pennucci}, T., {Ransom}, S.~M., {Roberts}, M.~S.~E., \&
  {Hessels}, J.~W.~T. 2010, \nat, 467, 1081

\bibitem[{{Gehrels} {et~al.}(2004){Gehrels}, {Chincarini}, {Giommi}, {Mason},
  {Nousek}, {Wells}, {White}, {Barthelmy}, {Burrows}, {Cominsky}, {Hurley},
  {Marshall}, {M{\'e}sz{\'a}ros}, {Roming}, {Angelini}, {Barbier}, {Belloni},
  {Campana}, {Caraveo}, {Chester}, {Citterio}, {Cline}, {Cropper}, {Cummings},
  {Dean}, {Feigelson}, {Fenimore}, {Frail}, {Fruchter}, {Garmire}, {Gendreau},
  {Ghisellini}, {Greiner}, {Hill}, {Hunsberger}, {Krimm}, {Kulkarni}, {Kumar},
  {Lebrun}, {Lloyd-Ronning}, {Markwardt}, {Mattson}, {Mushotzky}, {Norris},
  {Osborne}, {Paczynski}, {Palmer}, {Park}, {Parsons}, {Paul}, {Rees},
  {Reynolds}, {Rhoads}, {Sasseen}, {Schaefer}, {Short}, {Smale}, {Smith},
  {Stella}, {Tagliaferri}, {Takahashi}, {Tashiro}, {Townsley}, {Tueller},
  {Turner}, {Vietri}, {Voges}, {Ward}, {Willingale}, {Zerbi}, \&
  {Zhang}}]{Swift}
{Gehrels}, N., {Chincarini}, G., {Giommi}, P., {et~al.} 2004, \apj, 611, 1005

\bibitem[{{Ghirlanda} {et~al.}(2004){Ghirlanda}, {Ghisellini}, \&
  {Lazzati}}]{Ghirlanda2004}
{Ghirlanda}, G., {Ghisellini}, G., \& {Lazzati}, D. 2004, \apj, 616, 331

\bibitem[{{Gompertz} {et~al.}(2014){Gompertz}, {O'Brien}, \&
  {Wynn}}]{Gompertz2014}
{Gompertz}, B.~P., {O'Brien}, P.~T., \& {Wynn}, G.~A. 2014, \mnras, 438, 240

\bibitem[{{Gompertz} {et~al.}(2013){Gompertz}, {O'Brien}, {Wynn}, \&
  {Rowlinson}}]{Gompertz2013}
{Gompertz}, B.~P., {O'Brien}, P.~T., {Wynn}, G.~A., \& {Rowlinson}, A. 2013,
  \mnras, 431, 1745

\bibitem[{{Harry} {et~al.}(2010)}]{Harry2010}
{Harry}, G.~M., {et~al.} 2010, Class. Quantum Grav., 27, 084006

\bibitem[{{Huang} {et~al.}(2002){Huang}, {Dai}, \& {Lu}}]{Huang2002}
{Huang}, Y.~F., {Dai}, Z.~G., \& {Lu}, T. 2002, \mnras, 332, 735

\bibitem[{{Lasota} {et~al.}(1996){Lasota}, {Haensel}, \&
  {Abramowicz}}]{Lasota1996}
{Lasota}, J.-P., {Haensel}, P., \& {Abramowicz}, M.~A. 1996, \apj, 456, 300

\bibitem[{{L{\"u}} {et~al.}(2015){L{\"u}}, {Zhang}, {Lei}, {Li}, \&
  {Lasky}}]{Lu2015}
{L{\"u}}, H.-J., {Zhang}, B., {Lei}, W.-H., {Li}, Y., \& {Lasky}, P.~D. 2015,
  \apj, 805, 89

\bibitem[{{Metzger} \& {Berger}(2012)}]{Metzger2012}
{Metzger}, B.~D., \& {Berger}, E. 2012, \apj, 746, 48

\bibitem[{{Metzger} \& {Piro}(2014)}]{Metzger2014}
{Metzger}, B.~D., \& {Piro}, A.~L. 2014, \mnras, 439, 3916

\bibitem[{{Metzger} {et~al.}(2008){Metzger}, {Quataert}, \&
  {Thompson}}]{Metzger2008a}
{Metzger}, B.~D., {Quataert}, E., \& {Thompson}, T.~A. 2008, \mnras, 385, 1455

\bibitem[{{P{\'e}langeon} {et~al.}(2008){P{\'e}langeon}, {Atteia}, {Nakagawa},
  {Hurley}, {Yoshida}, {Vanderspek}, {Suzuki}, {Kawai}, {Pizzichini},
  {Bo{\"e}r}, {Braga}, {Crew}, {Donaghy}, {Dezalay}, {Doty}, {Fenimore},
  {Galassi}, {Graziani}, {Jernigan}, {Lamb}, {Levine}, {Manchanda}, {Martel},
  {Matsuoka}, {Olive}, {Prigozhin}, {Ricker}, {Sakamoto}, {Shirasaki},
  {Sugita}, {Takagishi}, {Tamagawa}, {Villasenor}, {Woosley}, \&
  {Yamauchi}}]{Pelangeon2008}
{P{\'e}langeon}, A., {Atteia}, J.-L., {Nakagawa}, Y.~E., {et~al.} 2008, \aap,
  491, 157

\bibitem[{{Pian} {et~al.}(2006){Pian}, {Mazzali}, {Masetti}, {Ferrero},
  {Klose}, {Palazzi}, {Ramirez-Ruiz}, {Woosley}, {Kouveliotou}, {Deng},
  {Filippenko}, {Foley}, {Fynbo}, {Kann}, {Li}, {Hjorth}, {Nomoto}, {Patat},
  {Sauer}, {Sollerman}, {Vreeswijk}, {Guenther}, {Levan}, {O'Brien}, {Tanvir},
  {Wijers}, {Dumas}, {Hainaut}, {Wong}, {Baade}, {Wang}, {Amati}, {Cappellaro},
  {Castro-Tirado}, {Ellison}, {Frontera}, {Fruchter}, {Greiner}, {Kawabata},
  {Ledoux}, {Maeda}, {M{\o}ller}, {Nicastro}, {Rol}, \& {Starling}}]{Pian2006}
{Pian}, E., {Mazzali}, P.~A., {Masetti}, N., {et~al.} 2006, \nat, 442, 1011

\bibitem[{{Planck Collaboration} {et~al.}(2015){Planck Collaboration}, {Ade},
  {Aghanim}, {Arnaud}, {Ashdown}, {Aumont}, {Baccigalupi}, {Banday},
  {Barreiro}, {Bartlett}, \& et~al.}]{Planck2015}
{Planck Collaboration}, {Ade}, P.~A.~R., {Aghanim}, N., {et~al.} 2015, ArXiv
  e-prints, arXiv:1502.01589

\bibitem[{{Ricker} {et~al.}(2003){Ricker}, {Atteia}, {Crew}, {Doty},
  {Fenimore}, {Galassi}, {Graziani}, {Hurley}, {Jernigan}, {Kawai}, {Lamb},
  {Matsuoka}, {Pizzichini}, {Shirasaki}, {Tamagawa}, {Vanderspek}, {Vedrenne},
  {Villasenor}, {Woosley}, \& {Yoshida}}]{Ricker2003}
{Ricker}, G.~R., {Atteia}, J.-L., {Crew}, G.~B., {et~al.} 2003, in American
  Institute of Physics Conference Series, Vol. 662, Gamma-Ray Burst and
  Afterglow Astronomy 2001: A Workshop Celebrating the First Year of the HETE
  Mission, ed. G.~R. {Ricker} \& R.~K. {Vanderspek}, 3--16

\bibitem[{{Rowlinson} {et~al.}(2013){Rowlinson}, {O'Brien}, {Metzger},
  {Tanvir}, \& {Levan}}]{Rowlinson2013}
{Rowlinson}, A., {O'Brien}, P.~T., {Metzger}, B.~D., {Tanvir}, N.~R., \&
  {Levan}, A.~J. 2013, \mnras, 430, 1061

\bibitem[{{Sakamoto} {et~al.}(2004){Sakamoto}, {Lamb}, {Graziani}, {Donaghy},
  {Suzuki}, {Ricker}, {Atteia}, {Kawai}, {Yoshida}, {Shirasaki}, {Tamagawa},
  {Torii}, {Matsuoka}, {Fenimore}, {Galassi}, {Tavenner}, {Doty}, {Vanderspek},
  {Crew}, {Villasenor}, {Butler}, {Prigozhin}, {Jernigan}, {Barraud}, {Boer},
  {Dezalay}, {Olive}, {Hurley}, {Levine}, {Monnelly}, {Martel}, {Morgan},
  {Woosley}, {Cline}, {Braga}, {Manchanda}, {Pizzichini}, {Takagishi}, \&
  {Yamauchi}}]{XRF020903}
{Sakamoto}, T., {Lamb}, D.~Q., {Graziani}, C., {et~al.} 2004, \apj, 602, 875

\bibitem[{{Sakamoto} {et~al.}(2005){Sakamoto}, {Lamb}, {Kawai}, {Yoshida},
  {Graziani}, {Fenimore}, {Donaghy}, {Matsuoka}, {Suzuki}, {Ricker}, {Atteia},
  {Shirasaki}, {Tamagawa}, {Torii}, {Galassi}, {Doty}, {Vanderspek}, {Crew},
  {Villasenor}, {Butler}, {Prigozhin}, {Jernigan}, {Barraud}, {Boer},
  {Dezalay}, {Olive}, {Hurley}, {Levine}, {Monnelly}, {Martel}, {Morgan},
  {Woosley}, {Cline}, {Braga}, {Manchanda}, {Pizzichini}, {Takagishi}, \&
  {Yamauchi}}]{Sakamoto2005}
{Sakamoto}, T., {Lamb}, D.~Q., {Kawai}, N., {et~al.} 2005, \apj, 629, 311

\bibitem[{{Siegel} \& {Ciolfi}(2016{\natexlab{a}})}]{Siegel2016a}
{Siegel}, D.~M., \& {Ciolfi}, R. 2016{\natexlab{a}}, \apj, 819, 14

\bibitem[{{Siegel} \& {Ciolfi}(2016{\natexlab{b}})}]{Siegel2016b}
---. 2016{\natexlab{b}}, \apj, 819, 15

\bibitem[{{Soderberg} {et~al.}(2004){Soderberg}, {Kulkarni}, {Berger}, {Fox},
  {Price}, {Yost}, {Hunt}, {Frail}, {Walker}, {Hamuy}, {Shectman}, {Halpern},
  \& {Mirabal}}]{Soderberg2004}
{Soderberg}, A.~M., {Kulkarni}, S.~R., {Berger}, E., {et~al.} 2004, \apj, 606,
  994

\bibitem[{{Soderberg} {et~al.}(2005){Soderberg}, {Kulkarni}, {Fox}, {Berger},
  {Price}, {Cenko}, {Howell}, {Gal-Yam}, {Leonard}, {Frail}, {Moon},
  {Chevalier}, {Hamuy}, {Hurley}, {Kelson}, {Koviak}, {Krzeminski}, {Kumar},
  {MacFadyen}, {McCarthy}, {Park}, {Peterson}, {Phillips}, {Rauch}, {Roth},
  {Schmidt}, \& {Shectman}}]{Soderberg2005}
{Soderberg}, A.~M., {Kulkarni}, S.~R., {Fox}, D.~B., {et~al.} 2005, \apj, 627,
  877

\bibitem[{{Urata} {et~al.}(2015){Urata}, {Huang}, {Yamazaki}, \&
  {Sakamoto}}]{Urata2015}
{Urata}, Y., {Huang}, K., {Yamazaki}, R., \& {Sakamoto}, T. 2015, \apj, 806,
  222

\bibitem[{{Yamazaki} {et~al.}(2002){Yamazaki}, {Ioka}, \&
  {Nakamura}}]{Yamazaki2002}
{Yamazaki}, R., {Ioka}, K., \& {Nakamura}, T. 2002, \apjl, 571, L31

\bibitem[{{Yu} {et~al.}(2013){Yu}, {Zhang}, \& {Gao}}]{Yu2013}
{Yu}, Y.-W., {Zhang}, B., \& {Gao}, H. 2013, \apjl, 776, L40

\bibitem[{{Zhang} {et~al.}(2004){Zhang}, {Dai}, {Lloyd-Ronning}, \&
  {M{\'e}sz{\'a}ros}}]{Zhang2004}
{Zhang}, B., {Dai}, X., {Lloyd-Ronning}, N.~M., \& {M{\'e}sz{\'a}ros}, P. 2004,
  \apjl, 601, L119

\bibitem[{{Zhang} \& {M{\'e}sz{\'a}ros}(2001)}]{Zhang2001}
{Zhang}, B., \& {M{\'e}sz{\'a}ros}, P. 2001, \apjl, 552, L35

\end{thebibliography}

\end{document}